# Understanding and improving social factors in education: a computational social science approach[1]

Nabeel Gillani[1] (ORCID), Rebecca Eynon[2] (ORCID)
[1]College of Arts, Media and Design and D'Amore-McKim School of Business, Northeastern University, Boston, MA, USA
[2]Oxford Internet Institute and Department of Education, University of Oxford, Oxford, UK

**Abstract**
Over the past decade, an explosion in the availability of education-related datasets has enabled new computational research in education.  Much of this work has investigated digital traces of online learners in order to better understand and optimize their cognitive learning processes.  Yet cognitive learning on digital platforms does not equal education.  Instead, education is an inherently social, cultural, economic, and political process manifesting in physical spaces, and educational outcomes are influenced by many factors that precede and shape the cognitive learning process.  Many of these are social factors like children's connections to schools (including teachers, counselors, and role models), parents and families, and the broader neighborhoods in which they live.  In this article, we briefly discuss recent studies of learning through large-scale digital platforms, but largely focus on those exploring sociological aspects of education.  We believe computational social scientists can creatively advance this emerging research frontier—and in doing so, help facilitate more equitable educational and life outcomes.

## Introduction

The journalist Thomas Friedman's famous declaration of 2012 as "The Year of the MOOC" (Friedman, 2013) heralded the beginning of a new era of education.  MOOCs—or massive, open, online courses—drew enthusiasm and optimism from a wide audience as potential enablers of more equitable global access to quality education.  In parallel, they sparked a new wave of computational research in education.  Emerging platforms and the large datasets they created inspired researchers to analyze how students engage with lectures and quizzes online (Kizilcec et al, 2013; Breslow et al., 2013); try and predict who is most likely to drop out of courses (Kloft et al., 2014); experiment with new methods for sequencing learning content (Zhao et al., 2018); and even deploy interventions designed to improve course completion rates (Kizilcec et al., 2017; Kizilcec et al., 2020).  We, too, were among these researchers, analyzing patterns of engagement in MOOC discussion forums to better understand the nature of communication and social engagement in these spaces (Gillani & Eynon, 2014; Gillani et al., 2014; Eynon et al., 2016).  Yet almost as quickly as they rose to

---

prominence, it became clear that MOOCs would not be a silver bullet for addressing disparities in educational access and outcomes, or 'disrupt' higher education as many believed: with the exception of several notable cases (Cadwalladr, 2012), most early MOOC participants were well-educated adults, hailing mostly from developed countries (Emanuel, 2013).

MOOCs, and digital learning environments more broadly, have helped shed light on learner behaviors and patterns that may have previously been difficult—if not impossible—to measure.  For example, after initial school closures due to the COVID-19 pandemic, data from digital learning platforms helped reveal how students in the US from lower-income neighborhoods were engaging much less with academic content than those in more affluent areas (Chetty et al., 2020).  Yet understanding learning processes through data from digital platforms hardly tells us everything we need in order to improve educational access and outcomes.  Across the world, there are still tremendous global achievement gaps (Graetz et al., 2020)—gaps that persist even within specific developed countries like the US, stemming from a myriad of factors like continued racial and income segregation in schools (Reardon & Owens); racial and gender biases among some teachers and other education leaders (Starck, 2020); and, broadly speaking, the crippling effects of poverty on nutrition (Walker, 2011), attention and cognition (Mani et al., 2013; McCloyd, 1998), self-confidence (Browman et al., 2019), and other out-of-school factors that impact the extent to which children are able to learn and grow.

In this light, it is clear that digital learning platforms, no matter how advanced, will always be limited in the extent to which they can improve educational and life outcomes for all students—especially those experiencing various structural disadvantages like poverty and racism.  So, too, will be the potential impact of computational research as a whole—even if such research becomes more solutions-oriented, as some have called for (Watts, 2017).  Yet the collection and discovery of new education-related datasets, combined with advances in computational methods spanning exploratory data analyses, machine learning, social network analysis, and other approaches offer promise in equipping researchers across disciplines with new tools to ask questions that can surface knowledge about educational processes and systems in ways that were previously difficult to imagine.

This promise has motivated several research efforts in the past one to two decades, parallel to the interest in digital learning platforms, to explore education-related datasets using a myriad of computational approaches.  Journals have hosted special issues on "Educational Data Science" (McFarland et al., 2021) and "Educational Research in a New Data Environment" (Reardon & Stuart, 2019), featuring research hailing from both social scientists who are increasingly leveraging computational methods in their work, and computer/data scientists with an interest in the social sciences.  As these and other related articles highlight, researchers are using advances in natural language processing to identify gender biases in textbooks (Lucy et al., 2020); social network analysis to design effective anti-bullying interventions (Paluck et al., 2016); and quasi-experimental methods to infer the effectiveness of teachers (Chetty et al., 2014) and guidance counselors (Mulhern, 2020), to name a few.  Instead of confining themselves to digital learning platforms, these and other studies represent a growing body of work that seeks to use computational methods to explore issues germain to education systems and institutions as they are experienced every day, "in real life", by students and families.

In this chapter, we focus on research that explores one dimension of such education systems: the social factors that shape educational access and outcomes for children aged birth through (approximately) 18 years of age.  In particular, we discuss recent computational work exploring how schools, families, and neighborhoods shape children's educational and life outcomes from an early age.  Many researchers and practitioners agree that schools, neighborhoods, and families all operate on and affect children's educational trajectories in meaningful ways (Purpose Built Communities, 2019), but often debate the relative influence of each.  After briefly reviewing several studies, we discuss several directions of opportunity for future work, and how computational social scientists may creatively apply their unique disciplinary and methodological backgrounds to pursue them.

Before proceeding, we make three notes.  One: while we discuss digital learning platforms and the aforementioned social factors as two separate categories of research and practice, our purpose in doing so is not to create a false dichotomy.  Society and technology are interwoven (Selwyn, 2019), and education is no different.  Instead, we make this distinction largely to highlight the emerging body of work in the latter, and encourage computational social scientists with an interest in applications to education to consider investigating these social factors even when the datasets may not be as readily available or easy to capture compared to data generated from digital learning ecosystems.  Indeed, There is significantly more work to do to better conceptualise the relationships between education, digital technologies and society to facilitate meaningful social computational science in education.  Two: most of the examples we use are drawn from studies conducted in the US.  While many of the themes we discuss vis-a-vis the US are relevant in other countries, we also acknowledge the importance of more research specifically focused on, and conducted within, other international contexts—especially developing contexts, given that much of what works in the developed world cannot be force-fitted into developing countries (Irani et al., 2010).  Three: while some of the studies we highlight leverage large datasets and recent advances in machine learning and other data science techniques, several others use more traditional quantitative methods (like linear regression analyses used for program evaluation / causal inference) as their main methodological tools.  We include these different types of studies to contrast what is meant by 'computational', inviting readers to conceptualize a broad methodological landscape for conducting computational social science research in education.

# Computational approaches for understanding and improving learning: a learning science and analytics view

As noted earlier, the proliferation of digital learning environments is generating large-scale "digital trace" data describing how learners engage with online lectures, assignments, and other materials.  Much of the academic research exploring these questions has focused on understanding and optimizing "cognitive learning" processes—i.e., the processes through which students acquire knowledge or skills pertaining to specific academic topics (Mayer, 2012).  Furthermore, many of these studies have been conducted by researchers with backgrounds in computer science, data science, and learning or cognitive science.  While a full review of this vast literature is out of scope for this Chapter, below, we briefly highlight several studies across two broad categories—intelligent tutoring and learning analytics—to demonstrate different ways in which researchers are applying computational techniques to make sense of, and even shape, data in digital learning settings. Throughout these sections, we refer to "AI" (artificial intelligence) and machine learning; we refer those who may be unfamiliar with these terms or their broader applications and risks in education to Gillani et al., 2023. We note that many of the approaches described below may straddle the line between computational social science and computational cognitive science—especially when the focus is largely on optimizing the individual's learning process.

## Intelligent tutoring

Intelligent tutoring systems (ITS) are tools that seek to adapt to a student's learning style and state in order to help them learn content and build skills in a way that is uniquely suited to their needs. Given the ease-of-assessment for simple mathematics problems, many of these ITS have focused on helping students learn math, though there are also examples in other disciplines, e.g. language learning[2].

The "I" in ITS often has different definitions for different tools. For example, some ITS are machine learning-based systems that seek to infer a student's knowledge state based on which problems they answer correctly or incorrectly (Ritter, 2007). These systems then provide students with problems that are most likely to be at their "learning edge"—i.e. the problems they haven't yet answered that they are most likely to get correct, given their prior history of answers. Other ITS, like (Kelly et al., 2013), use simple rules or heuristics to determine if and when a student has mastered some concept (e.g. if they answer three or more of a particular type of question correctly in a row). Experimental evidence has largely shown ITS to be effective in increasing students' grades and test scores (Shank, 2019). Of course, grades and test scores offer only one (limited) view into student learning, and methodological challenges in evaluating the efficacy of ITS—e.g., "site selection bias" (Allcott, 2015)—may limit our ability to fully understand their impact on educational outcomes.

More recently, some researchers have argued that the real value of ITS may not lie in their problem recommendations, but instead, in what they can reveal about the granular misconceptions students harbor vis-a-vis course material in order to better inform and support how human educators teach (Baker, 2016). For example, in a recent paper, an intelligent tutoring algorithm that used deep neural networks to model students' knowledge states also produced a granular map of how different types of concepts and questions relate to one another. This map was a byproduct of which questions students answered correctly and incorrectly (Piech et al., 2015). Such interpretations shift computational learner modeling away from a cognition-optimization process to one that aims to scaffold teacher-student interactions through "learning analytics".

## Learning analytics

In addition to bootstrapping new ITS, the proliferation of data from digital learning environments has also inspired the development and use of "learning analytics" to improve teaching and learning practices (Gašević et al., 2015).

A large body of research over the past 10 years has illustrated the myriad of ways that methods from artificial intelligence (particularly machine learning) can be applied to extract insights from learner data.  For example, one of the first studies on digital trace data generated in MOOCs used unsupervised machine learning to infer a typology of participants based on which types of course activities they engage with, and for how long (Kizilcec et al., 2013).  Another study applied linear regression to system log files from a learning management system (LMS)—which captured data on usage frequency and system access patterns—to illustrate how more "regular" learning (proxied by how regularly a user logs into the LMS) positively predicts performance on a final test (Jo et al., 2014).  To illustrate the value of data generated by learners as a part of using intelligent tutoring system usage, (Xing & Goggins, 2015) build a machine learning model to detect when students are "going off task" based on platform usage.

---

[2] Duolingo.com and Busuu.com, for example.

These are but a few of the numerous learning analytics studies that currently exist.  There is, undoubtedly, great potential in using machine learning techniques to make sense of the vast amounts of data being generated in learning contexts; however, learning analytics as a discipline is still too nascent to make conclusive claims about how mining and analyzing learners' digital traces can enhance teaching and learning practices.  Some researchers have explicitly called this out, highlighting that there is still very little evidence on how learning analytics supports learning and teaching—and of the reported evidence, how very little of it shows negative effects, perhaps suggesting a skew in the research community towards reporting positive results (Ferguson & Clow, 2017).  Other researchers have cautioned against reducing learning analytics to "counting clicks", calling instead for an approach to analysis that is grounded in existing theories of learning—and hence, more likely to enhance learning outcomes (Gašević et al., 2015).  Finally, in many cases, there is still a large gap between applications of computational methods to learner trace data and to what extent these applications end up being useful to teachers and learners—inspiring researchers to define new roles like "educational data scientists" (Agasisti & Bowers, 2017) and "learning engineers" (Thille, 2016) to try and bridge these gaps in order to make the technical contributions better serve humans.

# Computational approaches for understanding and improving learning: a social factors view

While computer scientists have driven a large portion of the work behind the above-described computational approaches to analyzing data from learning platforms, much of the computational social science research on the role of schools, families, and neighborhoods in shaping children's educational and life outcomes has been generated by applied micro economists and sociologists.  We review several of these studies below.

## Schools

The local neighborhood has had an important role in the planning and development of many school systems across the world.  For example, in the US, locality-specific movements were central to enabling free primary and secondary schooling (Goldin & Katz, 2008).  Unfortunately, one enduring legacy of this primarily place-based movement to expand access to education is a continued relationship between neighborhood characteristics, especially racial demographics and household income, and academic achievement.  These relationships have resulted in achievement gaps where low-income children of color are significantly less likely to perform well in school when compared to their higher-income, White counterparts (Reardon et al., 2018).

To appreciate why these gaps matter, it is worth reflecting on why schools matter. Namely, what is the purpose of schooling? Economists are often interested in how teachers and schools impact intergenerational outcomes, like future earnings (Chetty et al., 2011), though some have also explored how education might also lead to greater happiness in adulthood (Oreopoulos & Salvanes, 2011). Through these lenses, the purpose of education is to help equip a child with the knowledge, skills, and attitudes needed to achieve these outcomes. However, the philosopher Biesta argues that the purpose of education is defined by our values, and thus, a lack of explicit articulation of these values in recent conversations about education and educational measurement makes such conversations incomplete. Without providing "an answer", Biesta offers a framework to help structure debates about the purpose of education: education is about "qualification", or equipping young people with skills; "socialization", or helping young people to become a part of a collective social, cultural, political order; and "subjectification", or helping individuals become independent and autonomous citizens (Biesta, 2009; Eynon, 2022).

Differing notions of the purpose of education (and, implicitly, the values that shape those perceived purposes) have fueled different measures of what constitutes quality schooling. One measure that has been shown to correlate with long-term outcomes is "effectiveness"—or how much a child learns and grows, over time, at their school (as opposed to snapshot measures like test scores alone, which are only weakly correlated with growth/effectiveness measures). Under these measures of effectiveness, "learning and growing" are usually defined in terms of changes in performance on standardized tests. While this is still an inherently limited measure—test scores do not capture the full breadth of a child's educational journey or outcomes—there has been evidence that students who are exposed to more effective educational settings are also more likely to attend college, earn more as adults, and less likely to have teen pregnancies (Chetty et al., 2014).

What makes schools effective? Ensuring adequate funding and resources for students, regardless of socioeconomic or demographic background, is of critical importance. However, a recent study analyzing data from several charter schools in New York City showed that even schools with higher per-pupil expenditures do not always improve learning outcomes (Fryer & Dobbie, 2013). This suggests there are other—perhaps more difficult-to-measure factors—that also matter. One answer is that effective schools have effective adults in them: teachers, counselors, and other staff who are well-suited to help students achieve their potential. There is a large literature investigating the development of "teacher value-added" models—i.e., data-driven, often quasi-experimental methods for computing the causal effect that teachers have on children's learning outcomes (Koretz, 2008). These models have helped identify the impact teachers have on both shorter-term measures (like test scores), but also, longer-term outcomes like those highlighted above. Teachers, of course, are one of many adults that students may be exposed to in school: guidance counselors, too, can have a pivotal impact on the life trajectories of children. A recent paper exploited the fact that many high schools assign students to guidance counselors based on the starting letter of the students' last names—an effectively random assignment—to identify the impact of counselors on students' academic achievement and college-going behaviors (Mulhern, 2020). The study found that effective counselors (based on their impact on students' performance in other years) had roughly the same impact on students' academic outcomes as effective teachers—a surprising insight given that counselors in US public schools often serve an order of magnitude more students (hundreds) than teachers do.

Beyond their effectiveness in delivering instruction or offering guidance on education-related matters, teachers and counselors often play an important role in the lives of students by serving as role models and mentors.  A recent study leveraged a range of regression specifications to identify positive relationships between a child having at least one in-school mentor and their future academic achievement (Kraft et al., 2021).  A prior study exploiting random assignment of students to classrooms found that Black students matched to same-race teachers were more likely to graduate high school and enroll in college (Gershenson et al., 2018)—perhaps due to students' abilities to see themselves and their life experiences reflected in these adult stakeholders.  The counselors' study discussed above found similar relationships.  Even when such opportunities for students to recognize their own unique background and circumstances in adult role-model and mentor-like figures are short-lived, they can potentially have important effects.  For example (Riley, 2019) showed that screening a movie depicting a strong, relevant female role model could lead students in Uganda—particularly lower-performing females—to achieve higher scores on their final exams in Math.  Another study demonstrated how a motivational speech by former US First Lady Michelle Obama, delivered at a lower-performing school for girls in the UK, raised end-of-year performance on standardized tests (Burgess, 2016).

Schools, therefore, can serve as conduits to connecting children to effective teachers, counselors, role models, and mentors.  Still, the relationships between place, income, race, and education have limited the extent to which children and their families can access schools that afford such opportunities for students.  To combat this, cities around the world are increasingly offering "school choice": opportunities for families to select schools for their children beyond their immediate neighborhood contexts.  Increased choice, however, does not always translate into increased quality of education.  The resources, knowledge, and other assets families can tap into are likely to shape their ability to take advantage of such choices in schooling, with a reproduction of advantage often the outcome of such policy strategies (Ball et al., 2014).

## Parents and families

Parents and families influence education in multiple ways; here, we focus on one of those ways: school choice.  Family units, and parents in particular, are often tasked with selecting schools for children. Parents, however, are influenced by their own prior beliefs and other priorities for what constitutes a "good school" for their child.  For example, many parents' notions of school quality, and subsequent choices, are influenced by what they hear through their social networks vs. more formal indicators of quality (Ball & Vincent, 1998)—which may help explain why some parents' school preferences are more closely correlated to measures of "peer quality" at the school (measured by standardized test score achievement) than measures of school effectiveness (Abdulkadiroglu et al., 2019).  Furthermore, some parents are also influenced by school characteristics that do not necessarily directly indicate the quality of education the school offers—for example, the school's proximity to their home or other geographic characteristics (Bell, 2009).  These preferences may be a reflection of what parents value, or a symptom of various barriers they face, like limitations in their ability to transport students to/from certain schools; scheduling conflicts that prevent them from visiting schools to actively evaluate how well they are likely to serve their children; or other challenges that families living in poverty or facing other hardships may be disproportionately more likely to experience (Robertson et al., 2021).

Whether parents have an explicit choice in which schools they select for their children, or this choice is implicit by way of where families end up living, in recent years, parents have turned to school reviews platforms like GreatSchools.org for insights into the quality of potential schools their children may attend (Lovenheim & Walsh, 2017). A recent study, however, used quasi-experimental methods (treating the expansion of GreatSchools' ratings availability across the US as effectively random) to argue that more availability of school ratings information has actually exacerbated racial and income segregation (Hasan & Kumar, 2019). While perhaps counterintuitive, this is not surprising: school ratings websites often highly-weight snapshot-in-time test scores—which are notoriously correlated with the racial and income demographics of the schools (Barnum & LeMarr LeMee, 2019)—and hence indirectly drive parents to make decisions based on these factors instead of measures that more accurately capture school effectiveness. Our own research has underscored this possibility, leveraging recent advances in natural language processing to analyze how parents talk about schools in the reviews they post on GreatSchools and finding that such reviews reflect racial and socioeconomic differences in schools instead of their actual quality (i.e., effectiveness[3]) (Gillani et al., 2021). This raises questions about what kinds of information these sites contain and how this information might be repositioned, or changed altogether, to minimize unintended and potentially-harmful consequences.

While more information can sometimes exacerbate inequalities in education, in other cases, it can reduce "information frictions" and knowledge gaps parents might face when seeking to support their children's education. Here, too, computational methods—combined with behavioral science-inspired interventions—have played an important role in reducing such frictions (Bergman, 2019). Many of these studies fall beyond the realm of school choice and seek to engage parents as active sources of support in their children's educational experiences. For example, (Bergman & Chen, 2019) designed an SMS-based system to inform parents about their children's class absences and performance, finding that doing so significantly reduced the rate of course failures. Focusing on lower-income parents of prekindergarten children, (York and Loeb, 2014) ran an 8-month-long text messaging campaign in which they shared information highlighting the importance of literacy as well as suggestions for small-scale literacy-related activities that could be done around the home. They found a significant increase in parental engagement and literacy levels after children started school. In addition to their results, what's particularly notable about these studies is that they move from observational analyses to intervention design and evaluation. As we discuss later in the chapter, computational social scientists may be particularly well-suited to help bridge existing gaps between data analysis and the design/evaluation of interventions that seek to improve educational outcomes.

## Neighborhoods

There is an active debate in the social sciences about how much neighborhoods—versus families or schools—influence a child's future outcomes. Some scholars argue that a child's family has the greatest impact on their development, especially at an early age (Heckman, 2006), in part, by facilitating the development of character and other metacognitive traits like motivation (Heckman & Kautz, 2013).  Referencing the Perry Preschool Project, a famous early childhood intervention administered in the 1960s that included home visits to help strengthen parents' abilities to support their children's learning (Schweinhart et al., 1993), these scholars note how some of the intervention's greatest long-term effects (e.g. higher earnings, lower crime rates) accrued to those who lived in the worst neighborhoods.  They use this to argue for "the relative unimportance of ZIP codes in explaining the observed intergenerational program effects on the children of Perry participants" (Heckman & Karapakula, 2019).

---

[3] Though, in late 2020, GreatSchools overhauled its ratings methodology in an effort to address these issues, downweighting test scores and upweighting measures of student growth and how well schools serve students from different backgrounds: https://www.greatschools.org/gk/articles/why-we-changed-our-ratings/.

Others argue that schools, more than neighborhoods, shape a child's future trajectory.  E.g., Fryer & Katz (2013) use quasi-experimental evidence documenting changes in children's neighborhood and school quality—comparing findings from the randomized experimental voucher-based "Moving to Opportunity" (MTO) study (Katz et al., 2001) to academic gains of those who received lottery admissions to a high-quality charter school in New York City (Dobbie & Fryer, 2009)—to suggest high-quality schools affect a child's long-term academic achievement and earnings more than better neighborhoods.

Yet recent experimental and quasi-experimental analyses have revealed the large impact a child's neighborhood can have on their academics and future earnings.  These studies critically note that how much a neighborhood shapes a child's future is, on average, a function of how much time the child actually spends growing up there.  For example, a few years after Fryer & Katz, 2013 was published, several researchers revisited the MTO study with two new lenses: i) longer-term earnings data for children (via tax returns) who participated in the study, and ii) an analysis of how future earnings for these children differed depending on how old they were at the time of the move.  The researchers found that when children moved to a lower-poverty neighborhood before the age of 13, their college attendance rates and future earnings increased.  On the other hand, older children experienced no such effects (Chetty et al., 2016).

Members of this research team have conducted several additional studies leveraging administrative (intergenerational tax and Census) records to quantify the impact a child's geography has on their future outcomes, like earnings.  For example, they analyze data describing millions of families who moved across commuting zones in the US to find that exposure to better neighborhoods leads children's future earnings to converge at a rate of 4% per year that they spend in that neighborhood to the future earnings of children who were already living there (Chetty & Hendren, 2015).  The authors leverage several econometric tools to control for potential confounders (like family effects) and identify this effect as the "causal effect of place" on a child's future outcomes.  While identifying the underlying mechanisms driving these effects will require more research, the authors find a variety of neighborhood characteristics that positively correlate with upward mobility, including: i) lower poverty rates, ii) more social capital, iii) higher-quality schools (as measured by higher test scores), and iv) more college graduates (Chetty et al., 2018a).

In a separate paper, researchers disaggregated the effects of neighborhoods on future outcomes by children's race and gender to find that black boys were much less likely to achieve upward mobility than white boys, even when they grew up in similar neighborhoods.  Interestingly, one of the features of neighborhoods with smaller gaps between black and white boys' future mobility rates was the presence of more black fathers per capita (not necessarily a child's own father within the household)—pointing to the possible impact, again, of role model figures on facilitating better outcomes (Chetty et al., 2018b).  Finally, a recent collection of follow-on studies leveraged an unprecedented scale of Facebook data (70 billion friendships across 20 million individuals) to shed new empirical light on a long-standing topic: the belief that social capital, defined as friendships across socioeconomic divides, has a causal effect on intergenerational upward mobility (Chetty et al., 2022a; Chetty et al., 2022b).  These new studies illustrate how digital trace data and administrative records, together, can help illuminate potential pathways to improved life outcomes for youth.

Thus, much like schools, neighborhoods may affect children through how they expose (or fail to expose) children to role models and mentors.  This possibility is underscored by another recent study, which highlighted how children who grow up in geographies with more inventors are more likely to become inventors themselves in the same technology classes that are prevalent in those geographies (Bell et al., 2018).  Thus, children growing up in the Bay Area near Silicon Valley are more likely to invent new software/computer technologies; children growing up in the Midwestern US are more likely to invent medical devices; etc.  The authors argue that such differences point to a causal impact of exposure to innovation: children who are exposed to innovation are more likely to become inventors themselves when they grow up.  Unfortunately, girls, children of color, and those living in low-income environments are often less likely to be exposed to inventors who share their backgrounds, reflecting longstanding structural inequity in society.

These recent findings suggest that boundaries between schools, families, and neighborhoods may actually be more fluid and interleaving than some researchers have previously suggested.  If we think of schools as a part of many children's neighborhood experiences, and generalize the influence a family has to the impact their social network (which includes family, but also neighbors, peers, role models and others) has—it is clear that neighborhoods constitute a "meta-mechanism" that influence children's futures through a range of possible channels that include schools and family-like figures.

## Looking ahead: exploring social factors in education with computational social science

As many of the citations so far suggest, much of the existing computational work exploring social factors in education has been conducted by applied microeconomists.  However, there is a need and opportunity to draw more sociologists and philosophers of education into these debates—particularly to offer new perspectives on theories of social justice and the values that influence how we think about the role and purpose of education in an ever-evolving social context.  Some have argued that rapidly-expanding datasets and new computational methods will bring about "the end of theory" (Anderson, 2008), yet the creativity and nuance of many of the studies we've discussed in this chapter should make it clear that asking meaningful questions will require much more than data mining alone. Computational social scientists can partner with sociologists and philosophers to incorporate their rich theories to inform new empirical questions, while also illuminating how empirical inquiry can spark the development of novel theories and hypotheses (Blades, 2021).  Playing this intermediary role is sorely needed to more holistically investigate how social factors shape educational and life outcomes. Below, we outline some of the ways in which computational social scientists can help bridge these worlds.

## Measuring a broader set of inputs and outcomes

One of the inherent limitations in exploring social factors that shape children's educational and life outcomes is capturing the "right" data. One simple way to think of such data is in terms of inputs and outcomes. For example, throughout this Chapter, we've shared studies that explore relationships between inputs like students or teachers, and outcomes like academic achievements (as measured by performance on standardized tests). Despite the important role of social factors in shaping educational and life outcomes, few studies develop robust ways of measuring the strength of social ties (e.g. by capturing social network data between participants[4]), or other related measures that have been theoretically linked to improved outcomes for youth and families, like social capital (Chetty et al., 2018a; Putnam, 2000). Thus, looking ahead, can we develop more robust ways of capturing and incorporating social network information into existing models? For example, perhaps by using data from existing social networking platforms (a la recent efforts by Chetty et al., 2022a and Chetty, et al., 2022b)? Or by creating new methods for capturing the (evolving) social networks of students and critical peer and adult figures in their lives? Doing so may help offer more insights into the nature and kinds of social ties that are most associated with achievement. More ambitiously, future work could make even stronger connections with the long standing literature that makes far more visible the process and practices of schools—what is sometimes called the 'new sociology of education' (e.g. Young, 1972)—and with those focused on critical pedagogy (e.g. Friere, 1970).

Furthermore, in terms of outcomes, how might we broaden outcome measures beyond performance on tests to capture more about the impact social factors have on children's conceptions of self, feelings about their place and purpose in the world and in relation to others, or other aspects of living a quality of life that may be correlated with—but still distinct from—easier-to-measure, inherently quantitative factors like test scores (or even longer-term measures like future income)? These are difficult questions that will benefit from fresh, creative thinking from computational social scientists interested in developing new input and outcome measures from existing and new datasets—but that can only be achieved through meaningful engagement with other academic communities that have different philosophical, conceptual and often methodological approaches.

## Starting with the questions, not methods

One of the benefits of established disciplines is that they often have built up, over decades or even centuries, established approaches and methods to different types of research questions. For example, linear regression models and their many variants are often the go-to choices for many applied microeconomists. While such methodological focus can produce efficiencies in data analysis and sharing, it can also—over time—constrain the types of questions that even seem plausible to ask. For example, linear regression is limited as a tool for natural language processing, where the words and phrases may have complex, non-linear dependencies in sentences; where the feature space may be much larger than the number of observations in the dataset itself (and hence, prone to overfitting); etc. In recognition of this, social scientists are developing a richer cadre of "Text as Data" methods (Gentzkow et al., 2019).

---

[4] There are a few notable exceptions, e.g. (Paluck et al., 2016).

By starting with the questions and turning to methods as needed to explore and answer such questions as deeply and richly as possible, computational social scientists can help draw attention to a wider range of topics on the education research agenda—including those that may not directly identify causal relationships (which is of great interest to many applied microeconomists) but still help yield valuable insights that inform downstream causal analyses, design-based research, etc. (Singer, 2019). A byproduct of such an approach may be a greater degree of methodological diversity applied to different research problems, and over time, an expansion of the questions researchers and practitioners even deem possible to ask.

## Unpacking the underlying mechanisms driving certain observed outcomes

Much of the applied microeconomics literature on neighborhood effects or teacher effectiveness outline how much neighborhoods or teachers can increase children's shorter and longer-term outcomes, but few unpack at a granular level *what it is* about neighborhoods, or teachers, or other social factors that make them more or less effective. Indeed, looking back through the history of the sociology of education, we see a similar criticism of early work from economists and quantitative researchers in the 1960s and 70s exploring questions of equity and schooling, many of whom treated schools and associated institutions as a black box (Weiss, 2016). It is naive to think data and computational methods alone will enable researchers to answer these questions more deeply in the coming years; a rich body of qualitative research will continue to be indispensable. This will encompass a continuation of research that is broadly considered more qualitative or quantitative in the traditional sense, but there may also be opportunities for novel computational work that spans across these boundaries. Nevertheless, there is still a critical role of computational methods in this pursuit. For example, might we use techniques from computer vision applied to historical corpora of neighborhood-level Google Street View images (a la Naik et al., 2017) to better understand environmental predictors, and eventually, causal factors responsible for different rates of upward mobility across neighborhoods? Can we access and merge datasets spanning notes and audio/video recordings describing teacher classroom observations (a la Kelly et al., 2018), students' written feedback, and other inputs to identify behaviors and practices that lead to larger gains in life outcomes for students? These and other questions may benefit from a fresh look from computational social scientists hailing from different disciplinary backgrounds. They will also require thoughtful considerations of data privacy and ethics, as new streams of data will bring with them new questions about how they should or should not be processed—and what their unintended consequences might be (Hakimi et al., 2021).

Unpacking mechanisms is not just about providing richer research about particular phenomena, it is also about meaningfully engaging with researchers that have very different ontological and epistemological worldviews to further refine and develop the conceptual frames often seen in education data science (Eynon, 2023). Furthermore, unpacking mechanisms requires challenging our own notions of access and "success" in education, and not treating education as a straightforwardly positive and neutral 'thing'. What we have seen from the discussion above is that both the learning analytics/sciences and social factors perspectives offer a strong focus on distributional issues related to education. In other words, there is a tendency to focus primarily on questions related to the resources required to ensure everyone has access to a good quality education (typically measured in quite narrow ways related to achievement on standardized tests) (Keddie, 2012). This may be, for example, through the use of more technology in the classroom, the use of grants or vouchers, using technology to provide more or better information to help with school choice, or trying to ensure that where a person can afford to live does not impact the quality of the school systems a young person can access.

These are all important issues. However, questions of equity and social justice—which are intrinsically linked to issues of quality education—have multiple dimensions (North, 2006). Beyond questions of distribution, we must also explore the cultural and political dimensions of injustice (Fraser, 2008). Specifically, questions of misrecognition (such as cultural domination and disrespect) that marginalised communities may experience at school, through, for example, the ways that young people are treated, and what is / what is not included in the curriculum (Power and Frandiji, 2010:388). And questions of misrepresentation may arise due to potential deficits of educational governance across school systems which make participation for some groups significantly challenging (Sayed, et al., 2020).

Relatedly, we can enhance awareness of how some interventions designed to make schooling more equal can inadvertently reinforce the status quo or cause harm to different groups. Research has shown, for example, how the common focus on distributive approaches to social justice in schools tends only to actually further stigmatise "the poor" rather than change economic structures to make society fairer (e.g. North, 2006; Leibowitz & Bozalek, 2016). Achieving equity and justice in schools (and indeed in society more broadly) requires attention to the complex interplay between economic, cultural and political factors (Fraser, 2008).

Computational social scientists may partner with sociologists, educationalists, and philosophers of education or familiarize themselves with related theories and methods, to bring a more nuanced understanding of the complex role that education plays in society (beyond delivering learning and qualification) and ways of theorising about social justice and education that would add to the existing research in this domain.

## Bridging analysis and design

As computational social scientists begin to more deeply unpack the underlying mechanisms responsible for observed effects, they will be uniquely suited to also help inform, and perhaps even lead, the design and development of interventions that seek to improve educational and life outcomes for youth. Currently, a wide range of researchers participate in the design and evaluation of education-related interventions. Often, these interventions come in the form of applied microeconomists running randomized control trials like those highlighted earlier or in more international settings (Gibson & Sautmann, 2021). They also come, however, in the form of computing and design researchers using both human-centered and participatory design methods to develop frameworks and systems that seek to serve different segments of the population (e.g. Costanza-Chock, 2020). Not all computational social scientists may wish to participate in design-related activities that pick up where analyses of secondary data (or even preliminary causal analyses) leave off, but their methodological and domain insights may lend them to fill an important "in-between" space that connects theory and methods from research domains to the practical challenges and considerations of intervention in field settings. Engaging in these in-between spaces may help create new fruitful directions for research, and also, meaningful social change—with due deference and humility in light of how different cultures and groups of people may desire (or not desire) such interventions (Irani et al., 2010).

### Expanding the audience of research and evaluating/improving "how it lands"

Learning analysts and engineers often see digital platform developers or certain groups of users (e.g. teachers and school administrators) as important audiences for their findings. Applied micro economists and sociologists may largely target policymakers—often at state and national levels—as their primary audiences. As computational social scientists select questions to work on, they also have an opportunity to think creatively about which audience(s) they wish to engage directly in dialogue with their work. Much like our point about methods above, letting the questions drive the thinking around which audiences are appropriate for the work—instead of assuming a priori that the audience must be a government policymaker, or platform designer, or some other known stakeholder—may help surface new ideas for people and systems that may have interest in learning about and building upon the research findings that computational social scientists help produce.

Simply picking an audience, however, does not guarantee positive social change; there are significant barriers to communicating findings from research and science. The COVID-19 pandemic has made this particularly clear (if it wasn't already before): for example, school-based mask mandates were a hotly contentious issue across school districts in the US (Cottle, 2021), and more generally, vaccine hesitancy levels remain high (Sreedhar & Gopal, 2021). The ways in which individuals, organizations, policymakers, and governments as a whole process and act upon scientific findings is a complex function of individual motivations and preferences; sociopolitical forces; notions of fairness and morality; and several other factors (Haidt, 2012). There are also more tactical factors at play, like time scarcity (Rogers & Lasky-Fink, 2020) and the ability of decision-makers to identify the relevance of existing research to their own contexts (Nakajima, 2021). There is, therefore, a tremendous need for research that explores not only which policies and practices help improve educational outcomes, but also, *how those practices can most effectively be communicated and positioned in order to increase the chances of their eventual implementation*. These communications-based research questions are certainly pertinent to education, but also other domains, and could benefit from the creative thinking and new questions that computational social scientists are well-positioned to put forth and test.

## Conclusion

In this Chapter, we have presented two perspectives on how computational social scientists might engage with research pertaining to education. We have focused more on the social factors that shape educational outcomes—namely, schools, families, and neighborhoods—given the impact such factors have on influencing educational access, experiences, and outcomes. We believe it is an exciting time for computational social scientists to apply their interstitial interests and skill sets to questions in education—and to creatively come up with new questions that we can't even imagine yet.

## Further readings

Interested readers may look to (Fischer et al., 2020) for a more in-depth review of the ways in which computational methods are being used to better understand data from digital learning environments. Readers who wish to learn more about using behavioral science to foster greater parental engagement in schools may look to (Bergman, 2019). Finally, for a broader view on trends in the use of data science and other computational methods in education, we refer readers to (McFarland et al., 2021) and (reardon & Stuart, 2019).